\documentclass[12pt,notitlepage]{article}

\usepackage{amsmath,amssymb}
\usepackage{graphicx}
\usepackage{cite}

\usepackage{color}

\def\ignore#1{{}}

\newcounter{sxn}

\newcounter{axn}

\newdimen\mybaselineskip
\newcommand{\beeq}{\begin{equation}}
\newcommand{\eneq}{\end{equation}}
\newcommand{\beqn}{\begin{eqnarray}}
\newcommand{\eeqn}{\end{eqnarray}}

\newcommand{\ba}{\begin{array}}
\newcommand{\ea}{\end{array}}

\newcommand{\be}{\begin{equation}}
\newcommand{\ee}{\end{equation}}
\newcommand{\bea}{\begin{eqnarray}}
\newcommand{\eea}{\end{eqnarray}}
\newcommand{\beal}{\setcounter{letter}{1} \begin{eqnarray}}
\newcommand{\eeal}{\addtocounter{equation}{1} \end{eqnarray}}

\newcommand{\larrow}{\,\,\,\,\hbox to 30pt{\rightarrowfill}
\,\,\,\,}
\newcommand{\slarrow}{\,\,\,\hbox to 20pt{\rightarrowfill}
\,\,\,}

\def\nn {\nonumber}

\def\la{\raise.16ex\hbox{$\langle$}\lower.16ex\hbox{}  }
\def\ra{\, \raise.16ex\hbox{$\rangle$}\lower.16ex\hbox{} }

\def\psibar{ \psi \kern-.65em\raise.6em\hbox{$-$} \lower.6em\hbox{} }
\def\psibarb{ \psi \kern-.65em\raise.6em\hbox{$-$}  }

\def\nn {\nonumber}

\begin{document}

\thispagestyle{empty}





\begin{center}  
{\Large \bf Effect of dark matter on galactic black hole ringdown waveforms and shadows}

\vspace{1cm}

{\bf  Ramin G.~Daghigh$^{1}$ and Gabor Kunstatter$^2$}
\end{center}

\centerline{\small \it $^1$ Natural Sciences Department, Metropolitan State University, Saint Paul, Minnesota, USA 55106}
\vskip 0 cm
\centerline{}

\centerline{\small \it $^2$ Physics Department, University of Winnipeg, Winnipeg, MB Canada R3B 2E9}
\vskip 0 cm
\centerline{} 

\vspace{1cm}

\begin{abstract}
We calculate the effect of dark matter on the ringdown waveform and shadow of supermassive black holes at the core of galaxies.  Our main focus is on the supermassive black hole at the core of M87, which is large enough to allow for viable observational data.   We compare the effects of a dark matter spike  to those expected from a galactic halo of the same mass.  Our calculation for the halo starts from the Hernquist density
function and assumes anisotropic pressure that is zero in the radial direction. The
resulting Tolman-Oppenheimer-Volkoff equations allow the corresponding metric 
to be obtained analytically in closed form. The geometry of the anisotropic dark matter spike is the same as that obtained in [{\it ApJ} {\bf 940} 33 (2022)] under the assumption of isotropy.
The effect of the spike is orders of magnitude more significant than the halo as long as the distribution scale of the latter is within a few orders of magnitude of the value expected from observations.  Our results indicate that the impact of the spike surrounding M87* on the ringdown waveform may in principle be detectable. Finally, we point out the somewhat surprising fact that existing Event Horizon Telescope observations of black hole shadows are within an order of magnitude from being able to detect, or rule out, the presence of a spike. 
\end{abstract}

\newpage

\section{Introduction}

Dark matter (DM) halos are known to surround most, if not all, supermassive black holes at the center of galaxies. It has been proposed that DM spikes exist at the core of galactic halos 
\cite{DMspikeformation1,GondoloSilk,DMspikeformation3, DMspikeformation4}.  Several papers \cite{DMspikeGW1,DMspikeGW11,DMspikeGW2,DMspikeGW3,DMspikeGW4,DMspikeGW5,DMspikeGW6,DMspikeGW66} have recently derived the  effects of DM spikes on the gravitational waves emitted from black holes, but until recently the prospect of detecting these waves from supermassive black holes has been remote. This has now changed due to recent experimental developments.

The North American Nanohertz Observatory for Gravitational Waves (NANOGrav)\cite{NANOGrav}, Parkes Pulsar Timing Array (PPTA)\cite{PPTA}, and European  Pulsar Timing Array (EPTA)\cite{EPTA} have found a common background noise at frequencies around $10^{-8}$ Hz.  The very recent observations of Hellings–Downs correlations by the Chinese Pulsar Timing Array (CPTA) collaboration \cite{CPTA}, EPTA collaboration \cite{EPTA1}, NANOGrav collaboration \cite{NANOGrav1} and PPTA collaboration \cite{PPTA1}  strongly suggest  a stochastic gravitational-wave origin of this background noise. Supermassive black hole binaries are among the main candidates to produce this background gravitational wave in the nHz range.  A recent paper\cite{DMspikeGW7}  has looked for signatures of DM spikes surrounding supermassive black holes in this data. Although these observations are still very far from being able to measure ringdown waveforms from individual supermassive black holes, it is nonetheless important to determine accurately the effect of the DM distributions on these waveforms.

Another important gravitational signature of  black holes, namely the radius of the photon ring, has recently been measured for M87 using the remarkable images of the central black hole's shadow obtained by the  Event Horizon Telescope \cite{EHT-M87}.  Moreover,  the authors of \cite{EHT-M87-PRIMO} used novel statistical methods to reduce the error in this measurement to approximately $1\%$, leading to the possibility that measurements of the photon ring might one day yield information about the DM distribution surrounding supermassive black holes.

The  shadow size of the galactic black hole Sagittarius A* (Sgr A*) in the presence of a DM spike was investigated in \cite{Nampalliwar}.\footnote{Similar calculations for the shadow size of black holes surrounded by DM halos were done in \cite{Halo-Jusufi1, Halo-Jusufi2}.}  In a more recent paper \cite{DK-spike}, the present authors studied the impact of the DM spike on ringdown waveforms for the supermassive black holes Sgr A* and M87*. Using an approach similar to that of \cite{DK-spike} in solving the Tolman-Oppenheimer-Volkoff (TOV) equations, the authors of \cite{Zhao-spike} calculated the quasinormal modes (QNMs) that dominate the gravitational ringdown process of perturbed black holes surrounded by DM spikes.  In these previous papers, the calculations were done in terms of the Schwarzschild time of the central black hole as measured by an observer located between the black hole event horizon and the inner radius of the spike. 

In the following, we calculate the effect of DM spikes on the ringdown waveforms and shadows of the supermassive black holes at the center of Milky Way and M87 galaxies. We emphasize the importance of expressing the results in the frame of an asymptotic stationary observer for obtaining the correct predictions\footnote{The authors are grateful to Andrei Frolov for illuminating discussions in this regard.}.   Using the Schwarzschild time within the inner radius of the spike effectively neglects an overall redshift factor that provides the dominant contribution. 
Ignoring this redshift, as done in \cite{DK-spike}, underestimates considerably the size of the effect. 

In addition, using the method developed in \cite{DK-spike} combined with the redshift, we construct the spacetime for a black hole surrounded by a DM halo in order to compare its impact to the spike case. To do this, we start with the Hernquist density profile\cite{Hernquist}, which describes the distribution of DM halos in elliptical galaxies such as M87.

Two possible assumptions about the DM pressure have been used to reduce the problem of determining the spacetime geometry  to that of solving the standard TOV equations. One of these (see \cite{Nampalliwar} and \cite{DK-spike}) is the assumption of isotropy. Isotropic pressure is however suspect in the case of non-interacting dust since near the photon sphere, where the motion is highly relativistic, non-zero radial pressure necessarily implies a flow of matter into the black hole and renders the solution non-static\footnote{We are greatful to Don Page for bringing this argument to our attention.}. The assumption of isotropy can nonetheless be viable in certain regions of the dark matter halo \cite{Navarro, Ludlow} and in certain scenarios such as self-interacting dark matter spikes\cite{Saxton}.
The second, more physical, assumption is the anisotropic DM with zero radial pressure and non-zero uniform pressure tangential to spheres of constant radius. We will show in the following that the TOV equations imply  the same geometry to a high degree of accuracy for both assumptions when the spike density profile is near the range previously determined by observation. 

In \cite{CardosoHalo}, Cardoso {\it et al.~}found an analytic solution for a Hernquist-like density profile for an anisotropic   DM halo (with zero radial pressure) surrounding a black hole\footnote{For spacetime solutions for other DM halo density profiles, see \cite{KonoplyaHalo}}. In their solution, Cardoso {\it et al.~}modify the Hernquist density function in order to impose the boundary condition $g_{tt}(r\to\infty)\to1$, where $g_{tt}(r)$ is the metric function.  While the two approaches are different,  the solution of \cite{CardosoHalo} turns out to be a close match to the solution we find without altering the Hernquist density function.


In the context of shadows,  we confirm that the presence of a DM spike/halo increases  the size of the black hole shadow\cite{Nampalliwar,DK-spike}.  We point out that for the spike density estimated in \cite{DK-spike}\footnote{We note a typo in Eqs.\ (30) and (31) of \cite{DK-spike}, where the authors estimate the DM spike density based on observational data.  More specifically, the term $\left( 1+r_0/r \right)$ should be $\left( 1+r/r_0 \right)$.  The numerical results, however, are correct.}, based on the observational data provided in \cite{M87-observation}, the increase of the shadow size due to a DM spike can be of the order of $0.1\%$. This calculation has taken on new significance given the recent results of \cite{EHT-M87-PRIMO} for M87. The increase in the shadow size is just one order of magnitude less than the proposed error bars of \cite{EHT-M87-PRIMO}.
	

The paper is organized as follows. 
Section \ref{Sec:TOV} reviews the general derivation of the metric for both the isotropic and anisotropic cases and describes the subtleties associated with the choice of reference frame. Section \ref{Sec:WE} derives the wave equation, which is covariant under the corresponding change of frame. The resulting solution for the waveform is a scalar whose shape depends on the choice of coordinates. Section \ref{Sec:Numerics} presents a discussion of galactic black holes with DM spikes and their associated ringdown waveforms.  
Section \ref{Sec:DMH} derives the metric for a dark matter halo described by the Hernquist profile. The effects of the halo on the QNM potential, due to the redshift, are then compared to those expected from a DM spike of the same mass.
Section \ref{Sec:Shadow} reviews the shadow calculations and conclusions are given in Section \ref{Sec:Conclusion}. Appendix \ref{Sec:Appendix} is devoted to deriving an approximate expression for  the discontinuities in the QNM potential that necessarily exist at the boundaries of a DM spike. The discontinuities are proportional to the change in matter density at the boundaries and negligibly small for the proposed spike profiles.

\section{Deriving the Spacetime Metric}
\label{Sec:TOV}

In order to calculate ringdown waveforms that might emerge from compact objects, one first needs to determine the static background metric  based on the conjectured matter content surrounding the object. We are thinking here of DM halos/spikes. This requires solving the TOV equations with appropriate boundary conditions.

One starts with the most general $4$-D spherically symmetric static metric (up to coordinate transformations)
\bea
ds^2&=&-A(r)dt^2 +B(r)^{-1}dr^2 + r^2 d\Omega^2 \nonumber\\
&=& -e^{\mu(r)}dt^2 +\left(1-\frac{2M(r)}{r}\right)^{-1}dr^2 + r^2 d\Omega^2,
\label{eq:GeneralMetric}
\eea
where we use  geometric units with $c=G=1$.  For asymptotically flat metrics, ${e^{\mu(r)}\to \text{constant}}$ at spatial infinity. This constant is normally chosen to be unity, thereby fixing the time coordinate to be the proper time of a stationary asymptotic observer.
This is the correct choice when comparing waves emitted by distant compact objects, despite the fact that it has not always been used in recent discussions of ringdown waveforms.

We now outline the derivation of the metric in the spacetime containing a general anisotropic, extended but finite, spherical shell of matter surrounding a black hole.  See \cite{DK-spike} for more details on the derivation.  We denote the inner radius of the shell by $r_\text{in}$ and the outer radius by $r_\text{out}$.    
Assume a  spherically symmetric anisotropic perfect fluid stress tensor for the shell of matter
\bea
T^\mu_\nu = \hbox{diag}(\rho(r), p_r(r), p_t(r), p_t(r))~,
\eea
where $p_r$ is the pressure in the radial direction and $p_t$ denotes the pressure tangential to a 	symmetric two sphere at fixed radius, $r$.
The resulting equations  in the shell region are
\bea
G_{tt}&=0& \implies \frac{dM(r)}{dr} = 4\pi r^2 \rho(r)
\label{eq:Gtt}\\
G_{rr}&=0& \implies \frac{d\mu(r)}{dr} = 2\frac{M(r)+4\pi r^3 p_r(r)}{r\left[r-2M(r)\right]}
\label{eq:Grr}\\
\nabla_\nu T^{r\nu} &=& 0 \implies \frac{dp_r(r)}{dr}=-[\rho(r)+p_r(r)]\frac{M(r)+4\pi r^3p(r)}{r\left[r-2M(r)\right]}+ \frac{2}{r}(p_t-p_r)~.
\label{eq:MomentumConservation}
\eea
Two different cases have been considered in the recent literature.  

The first case is the isotropic case, where $p_t(r)=p_r(r)=p(r)$. Equations (\ref{eq:Gtt}) and (\ref{eq:Grr}) remain unchanged, except that $p_r$ is replaced by $p(r)$ in the latter. The momentum conservation equation becomes
\bea
\nabla_\nu T^{r\nu} &=& 0 \implies \frac{dp(r)}{dr}=-[\rho(r)+p(r)]\frac{M(r)+4\pi r^3p(r)}{r\left[r-2M(r)\right]} ~.
\label{eq:MomentumConservationIsotropic}
\eea
Equations (\ref{eq:Gtt}), (\ref{eq:Grr}), and (\ref{eq:MomentumConservationIsotropic}) are the standard spherical TOV equations found in textbooks (see \cite{Carroll} for example). They are rather difficult to solve in general because of the presence of $p(r)$ in both (\ref{eq:Grr}) and (\ref{eq:MomentumConservationIsotropic}). 

The second case is the anisotropic case considered in \cite{KonoplyaHalo, CardosoHalo} with zero radial pressure ($p_r=0$).   Eq.~(\ref{eq:Gtt}) again remains unchanged, while the remaining equations simplify to
\bea
G_{rr}&=0& \implies \frac{d\mu(r)}{dr} = 2\frac{M(r)}{r\left[r-2M(r)\right]}
\label{eq:GrrNonIsotropic}\\
\nabla_\nu T^{r\nu}&=& 0 \implies 0=-\rho(r)\frac{M(r)}{r\left[r-2M(r)\right]}+ \frac{2}{r}p_t ~.
\label{eq:MomentumConservationNonIsotropic}
\eea
In this case, Eq.\ (\ref{eq:GrrNonIsotropic})  can be integrated directly while Eq.~(\ref{eq:MomentumConservationNonIsotropic}) can be solved algebraically once $M(r)$ is determined from (\ref{eq:Gtt}). Note that one does not need to know the transverse pressure to determine the metric.

In both cases there are three equations in four unknowns, either $[\mu(r), M(r), \rho(r), p(r)]$ or $[\mu(r), M(r), \rho(r), p_t(r)]$, which means they need to be supplemented by a fourth equation. Normally this is taken to be the equation of state relating the density to the pressure. In the cases we are interested in, such as DM halos/spikes, theoretical calculations \cite{GondoloSilk} predict a particular density profile for the shell. This provides the extra equation. Therefore, there is no freedom left to specify the equation of state.

To solve the isotropic case, we will assume for the simplicity that  the pressure term is negligible when solving for $\mu(r)$ in Eq.\ (\ref{eq:Grr}). The validity of this assumption, in the case of DM spikes, was proven in \cite{DK-spike} and is confirmed again here.  As long as  the pressure term is negligible  in Eq.\ (\ref{eq:Grr}), it should be clear that, for a given density function, there is no difference in geometry for the isotropic case and the anisotropic case. This is a direct consequence of the fact that in both cases the pressure is irrelevant to the determination of the geometry, albeit for very different physical reasons. 


Thus, in both cases, given the density profile, one can obtain the mass function by integrating Eq.\ (\ref{eq:Gtt}) from $r_\text{BH}$ to spatial infinity with the boundary condition that $M(r<r_\text{in})=M_\text{BH}$, where we use $r_\text{BH}$ and $M_\text{BH}$ to indicate the horizon radius and mass of the central black hole respectively.
Given the mass function obtained above, one  integrates Eq.\ (\ref{eq:Grr}) to get the second metric function $\mu(r)$.
\label{item:BCs}
Since the TOV equations are first order, each requires the specification of a single boundary condition in order to provide a unique solution. For Eq.\ (\ref{eq:Gtt}), the correct boundary condition is that $M(r< r_\text{in})=M_\text{BH}$. Equivalently, one can choose ${M(r\ge r_\text{out})=M_\text{total}=M_\text{BH}+M_\text{shell}}$, where $M_\text{shell}$ is the total mass of the shell. Since the tangential metric must be continuous at the boundaries for the surface geometry on the boundary  to be uniquely defined, there is no freedom in the choice of boundary conditions.
A potential subtlety  does arise in the choice of boundary condition for $g_{{ t}{ t}}(r)$ (and consequently for $\mu(r)$) in Eq.\ (\ref{eq:Grr}), which in turn determines the time coordinate. The obvious choice is that $g_{{t}{t}}(r)\to 1$ ($\mu(r)\to 0$) far from the black hole so that  ${ t}$ is the proper time of an asymptotic observer at rest relative to the black hole.  This is presumably the frame in which the measurements of the ringdown waveform and shadow are made.
On the other hand, it is useful to consider the ringdown waveforms caused by outgoing waves  that originate in the vacuum region $r<r_{\text{in}}$, where the initial data are specified.  Such outgoing boundary conditions are preferred for two reasons: first, they are in some sense more physical since the ringdown waveforms from the merger of binary black holes do originate from near the horizon. Secondly, numerical integrations of the wave equation yield better results (i.e. a larger number of reliable oscillations before errors set in)  than simulations that start from ingoing waves.
When starting with outgoing initial data, as done in \cite{DK-spike}, it seems natural to impose the boundary condition on the metric in the shell's interior, namely
\be
g_{tt}(r<r_\text{in})= 1- \frac{2M_{BH}}{r}\, ,
\label{eq:InteriorBC}
\ee
which is the Schwarzschild metric of the central black hole. 

Following \cite{DK-spike} we integrate outward starting at $r_\text{in}$ with the boundary condition (\ref{eq:InteriorBC}), which fixes the time coordinate $t$ to be the Schwarzschild time of the vacuum metric associated with the central black hole. In this case, $t$ is what would be the proper time of a stationary observer at infinity had the DM spike not been present. This is not the same as the actual proper time of an asymptotic observer when the spike is present. It is the latter proper time with respect to which measurements will be taken.  We can rewrite Eq.\  (\ref{eq:InteriorBC}) in terms of $\mu(r)$ as
	\beeq
	\mu(r<r_\text{in})=\ln \left(1-\frac{2 M_\text{BH}}{ r}\right),
	\label{}
	\eneq
where  clearly $\mu(r) \rightarrow 0$ as  $r \rightarrow \infty$.

The integration of Eq.~(\ref{eq:Grr}) proceeds to larger radii with the assumption that the metric function be continuous at both shell boundaries. 
This is required to ensure that the shell boundary has a well-defined geometry.\footnote{Note that $\mu(r)$ is not necessarily smooth at the boundaries.  This introduces discontinuities in the QNM potential, which can be shown to be small in the present context.  For details, see Appendix \ref{Sec:Appendix}.}  
In the shell region $ r_{\text{in}}\le r < r_\text{out}$, we find the metric function to be 
	\beeq
	\mu(r)= \ln \left(1-\frac{2 M_\text{BH}}{ r_\text{in}}\right)+ \int_{r_\text{in}}^{r}  dr \frac{2 M(r)}{r(r-2M(r))}~.
	\label{eq-mu1}
	\eneq
To obtain a solution that is continuous at the outer radius of the shell, we extend the integral in Eq.~(\ref{eq-mu1}) past $r_\text{out}$ as follows:
	\beeq
	\mu(r \ge r_\text{out}) = \ln \left(1-\frac{2 M_\text{BH}}{ r_\text{in}}\right)+ \int_{r_\text{in}}^{r_\text{out}}  dr \frac{2 M(r)}{r(r-2M(r))}
	+\int_{r_\text{out}}^{r}  dr \frac{2 M_{\text{total}}}{r(r-2M_{\text{total}})} ~.
	\label{eq: muOut1}
	\eneq
	Note that the last integral above gives
	\beeq
	\int_{r_\text{out}}^{r}  dr \frac{2 M_{\text{total}}}{r(r-2M_{\text{total}})}= \ln \left(1-\frac{2 M_{total}}{ r}\right) -\ln \left(1-\frac{2 M_{total}}{ r_\text{out}}\right).
	\label{eq: muOut2}
	\eneq
	After combining  Eqs.\ (\ref{eq: muOut1}) and (\ref{eq: muOut2}), we arrive at
	\beeq
	\mu(r \ge r_\text{out}) = \ln \left(1-\frac{2 M_\text{total}}{ r}\right)+\ln(\mathcal C)~,
	\label{eq: muOut}
	\eneq
	where 
	\beeq
	{\mathcal C}=\frac{\left(1-\frac{2 M_\text{BH}}{r_{\text{in}}}\right)}{ \left(1-\frac{2 M_{total}}{ r_\text{out}}\right)}\exp{\int_{r_\text{in}}^{r_\text{out}}  dr \frac{M(r)}{r(r-2M(r))}}.
	\label{eq:b}
	\eneq 
	One thereby obtains a metric on the exterior that is asymptotically flat but for which $\mu(r)$ does not go to zero (i.e. $g_{tt}(r)\not\rightarrow 1$) at infinity: 
	\beeq
	ds^2 =- {\mathcal C}\left(1-\frac{2 M_\text{total}}{ r}\right) dt^2 +\left(1-\frac{2M_\text{total}}{r}\right)^{-1}dr^2 + r^2 d\Omega^2.
	\label{eq:MetricInsideTau}
	\eneq
The constant $\mathcal{C}\neq 1$ arises due to the presence of the matter shell so that $t$ is not the proper time of an asymptotic observer. 
It turns out that the constant $\mathcal{C}$ provides the dominant contribution to the modification of the ringdown waveform caused by the presence of the shell.
The relationship between the time $t$ in  (\ref{eq:MetricInsideTau}) and  the proper time $\tilde t$ of an asymptotic observer is
\bea
\tilde t= \sqrt{\mathcal{C}}t ~.
\eea
This is simply a redshift factor that is a consequence of the extra mass in the system provided by the shell. 
For positive definite shell densities $ \mathcal{C}$ is greater than one, so that in terms of the rescaled time $\tilde t$, the waveform appears stretched. 


\section{Wave Equation}
\label{Sec:WE}

In terms of  the Schwarzschild time $t$ of the vacuum metric of the central black hole, the QNM wave equation has the general form  
\beeq
\left( \frac{\partial^2}{\partial t^2}-\frac{\partial^2}{\partial r_*^2}+V(r(r_*))\right)\Psi(t, r_*)=0,
\label{eq:WaveInside}
\eneq  
where $r_*$ is the tortoise coordinate defined as
\beeq
dr_*=\frac{dr}{\sqrt{A(r)B(r)}}.
\label{eq:tortoise}
\eneq

For the particular case of massless scalar perturbations, the QNM potential is
\be
V(r)=A(r) \frac{l(l+1)}{r^2}+ \frac{1}{2r} \frac{d}{dr} \left[ A(r)B(r) \right]~.
\label{eq-scalarV}
\ee
With the given potential, one then solves (\ref{eq:WaveInside}) for the wave $\Psi(t,r_*)$   with the initial condition of an outgoing wave
\bea
\Psi(0,r_*)&=&f(r_*) \nonumber \\ 
\left. \partial_t \Psi(t,r_*)\right|_{t=0}&=&-\partial_{r_*}\Psi(0,r_*)~,
\label{eq:InitCond}
\eea
where $f(r_*)$ is usually taken to be a Gaussian function.  

As explained earlier, in order to find the waveform measured by an asymptotic observer, one can first determine the waveform in terms of the  Schwarzschild time $t$ of the interior geometry ($r<r_\text{in}$) by solving Eq.\ (\ref{eq:WaveInside}) with  initial conditions (\ref{eq:InitCond}) and then rescale the time coordinate to the proper time of an asymptotic observer, $\tilde{t}=\sqrt{\mathcal{C}} t$. Alternatively, one can do the entire calculation in terms of $\tilde t$.    In the latter case, one calculates ${\tilde \mu}(r)$ by integrating inward from $r_\text{out}$ with the boundary condition
\beeq
\tilde{\mu}(r\ge r_\text{out})=\ln \left(1-\frac{2 M_\text{total}}{ r}\right).
\label{}
\eneq
The integration of Eq.~(\ref{eq:Grr}) then proceeds to the shell region of $r_\text{in} \le r < r_\text{out}$, where
\beeq
\tilde{\mu}(r)= \ln \left(1-\frac{2 M_\text{total}}{ r_\text{out}}\right)+ \int_{r_\text{out}}^{r}  dr \frac{2 M(r)}{r(r-2M(r))}.
\label{eq-mu3}
\eneq
Combining Eqs. (\ref{eq-mu1}), (\ref{eq:b}), and (\ref{eq-mu3}), one finds that $\tilde{\mu}(r)=\mu(r) -\ln(\mathcal{C})$, which means 
\bea
\tilde{A}(r) =A(r)/\mathcal{C}. 
\label{eq-TildeA}
\eea 
Therefore, the conversion of the outward integration to an inward integration suitable for an asymptotic observer can be reduced to a simple rescaling.  
The tortoise coordinate for the asymptotic observer is
\beeq
d\tilde{r}_*=\frac{dr}{\sqrt{\tilde{A}(r)B(r)}}=\sqrt{\mathcal{C}} dr_*.
\label{eq:tortoise}
\eneq
In order to verify the covariance of the wave equation under time rescales, we note that
\be
\tilde{V}(r)=\tilde{A}(r) \frac{l(l+1)}{r^2}+ \frac{1}{2r} \frac{d}{dr} \left[ \tilde{A}(r)B(r) \right]=\frac{V(r)}{\mathcal{C}}.
\label{eq-scalar-tildeV}
\ee
Putting these transformations together we find, as expected, that
\beeq
\mathcal{C}\left( \frac{\partial^2}{\partial \tilde{t}^2}-\frac{\partial^2}{\partial \tilde{r}_*^2}+\tilde{V}(\tilde r_*)\right)\tilde\Psi(\tilde{t}, \tilde{r}_*)=0~,
\label{eq:WaveOutside}
\eneq  
where
\bea
\tilde{V}(\tilde r_*)=\frac{V(r(\tilde{r}_*))}{\mathcal{C}}.
\eea
Eq.~(\ref{eq:WaveOutside}) is of the same form as the wave equation (\ref{eq:WaveInside}).  This is all simply a manifestation of the fact that the field is a scalar, so that the wave equation is covariant and the solutions are related by
\bea
\tilde\Psi(\tilde t,\tilde{r}_*)=\Psi(t,r_*)=\Psi(\frac{1}{\sqrt{\mathcal{C}}} \tilde{t},\frac{1}{\sqrt{\mathcal{C}}}\tilde{r}_*)~.
\eea

In the frame of the asymptotic observer, the same initial conditions as in Eq.\ (\ref{eq:InitCond}) above take the form
	\bea
	\tilde\Psi(0,\tilde{r}_*)&=&{\tilde f}(\tilde{r}_*)= f(r_*(\tilde{r}_*))=f(\frac{1}{\sqrt{\mathcal{C}}}\tilde{r}_*) \nonumber \\
	\left.\sqrt{\mathcal{C}} \partial_{\tilde{t}} \tilde\Psi(\tilde{t},\tilde{r}_*)\right|_{\tilde{t}=0}&=&-\sqrt{\mathcal{C}}\partial_{\tilde{r}_*}\tilde\Psi(0,\tilde{r}_*) = \sqrt{\mathcal{C}}\partial_{\tilde{r}_*}f\left(\frac{1}{\sqrt{\mathcal C}}\tilde{r} \right)~.
	\eea
	Thus Gaussian initial data specified in terms of $r_*$ change the width by a factor of $\sqrt{\mathcal{C}}$ under the frame transformation. Of course, both forms correspond to the same initial data in terms of the areal radius. 
Although it is important to compare apples to apples with respect to initial data, numerical calculations are not sensitive to small changes to the initial wave due to the fact that black holes oscillate with natural vibrational modes (QNMs) that have unique frequencies and damping rates.

\section{Galactic Black Holes with Dark Matter Spikes}
\label{Sec:Numerics}

As we mentioned earlier, the results in \cite{DK-spike} are obtained in the Schwarzschild coordinates in the region $r<r_\text{in}$.   In this section, we examine the same waveforms as seen by an asymptotic stationary observer. 

The adiabatically formed DM spike, studied in \cite{DK-spike}, has a density profile\cite{GondoloSilk}
\begin{equation}
	\rho_{\text{DM}}^\text{sp}(r)\simeq \rho_{\text{sp}} \left( \frac{R_\text{sp}}{r} \right)^{\gamma_\text{sp}},
	\label{eq-SpikeDensity}
\end{equation}
were $\rho_\text{sp}$ and $R_{\text{sp}}$ are the density and radius, respectively, at the outer edge of the spike.  The expected value for $\rho_\text{sp}$, when $\gamma_\text{sp} = 7/3$, is taken to be  $8.00 \times 10^{-23}$ g cm$^{-3}$ for Sgr A (see Table 1 of \cite{DK-spike}) and $2.12 \times 10^{-23}$ g cm$^{-3}$ for M87 (see Table 2 of \cite{DK-spike}).  In this section, we present the results for these cases.  

\begin{figure}[th!]
	\begin{center}
		\includegraphics[height=5.3cm]{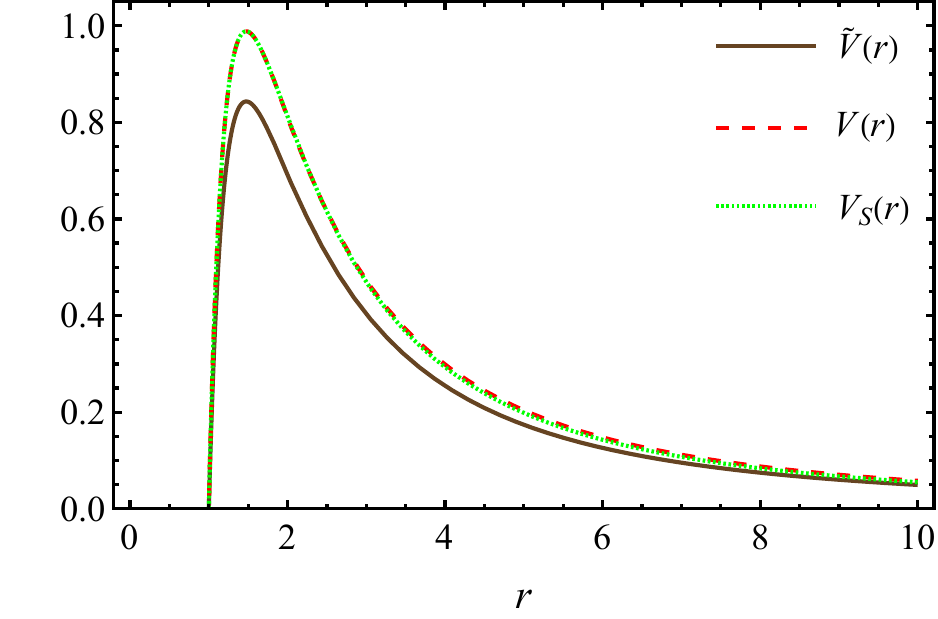}
	\end{center}
	\vspace{-0.7cm}
	\caption{\footnotesize Scalar QNM potential as a function of radial coordinate for $l=2$ and $\gamma_{\text{sp}} =7/3$ for Sgr A* with the mass of $M_\text{BH}=4.1\times 10^6 M_\odot$ surrounded by a DM spike. $V(r)$, in dashed red, is the potential corresponding to $t$  and $\tilde{V}(r)$, in solid brown, is the potential viewed by an asymptotic observer. We choose  $\rho_{\text{sp}}=6.7 \times 10^{-20}$ g cm$^{-3}$ ($\approx 840$ times the expected value), $r_\text{b}=2 r_{\text{BH}}$ and $R_{\text{sp}}=0.235$ kpc.  With these values, $\mathcal{C}=1.172$.  For comparison, we include the Schwarzschild potential, $V_S(r)$, in dotted green for $M_\text{BH}=4.1\times 10^6 M_\odot$. All our variables are expressed in terms of black hole parameters.}
	\label{fig-MW-potential}
\end{figure}
\begin{figure}[th!]
	\begin{center}
		\includegraphics[height=5.cm]{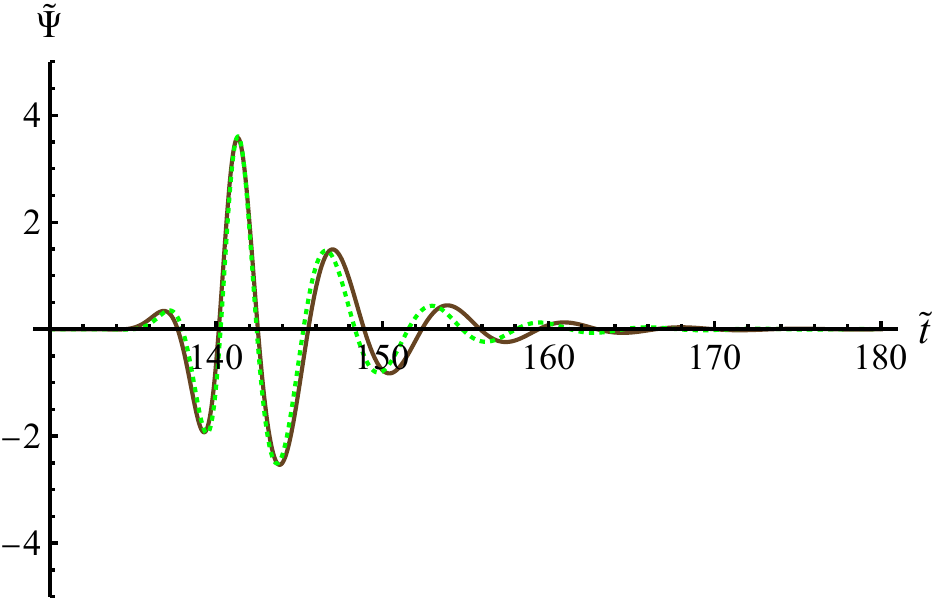}
		\includegraphics[height=5.cm]{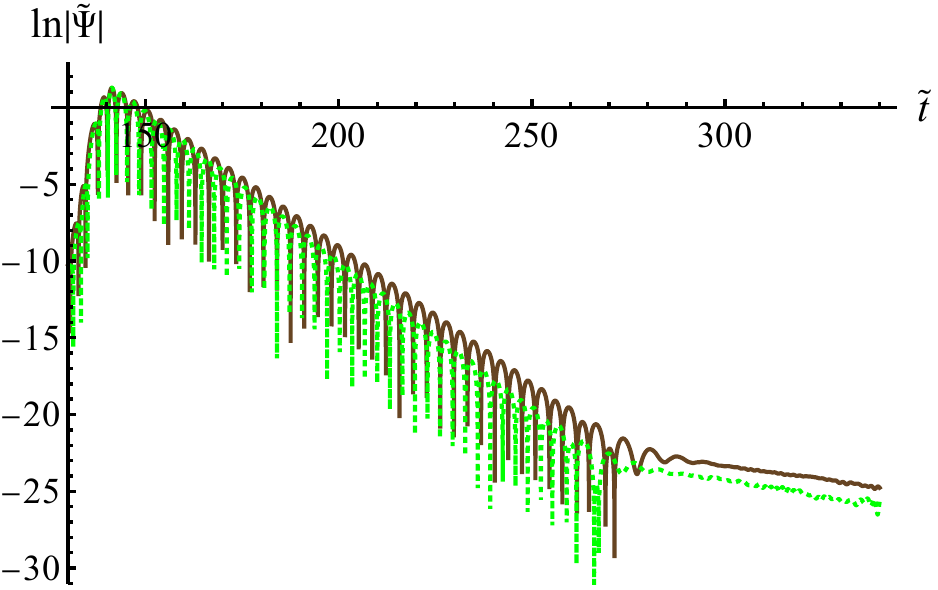}
	\end{center}
	\vspace{-0.7cm}
	\caption{\footnotesize In solid brown, ringdown waveform $\tilde{\Psi}$ (left) and $\ln |\tilde{\Psi}|$ (right) as a function of time $\tilde t$ for the potential, $\tilde{V}(r)$, shown in solid brown in Figure \ref{fig-MW-potential}.  For numerical purposes, we detect the waveform at $r=80 ~ r_{\text{BH}}$.  For comparison, we include the Schwarzschild ringdown waveform in dotted green. All our variables are expressed in terms of black hole parameters.}
	\label{fig-MWwave}
\end{figure}

\begin{figure}[th!]
	\begin{center}
		\includegraphics[height=5.3cm]{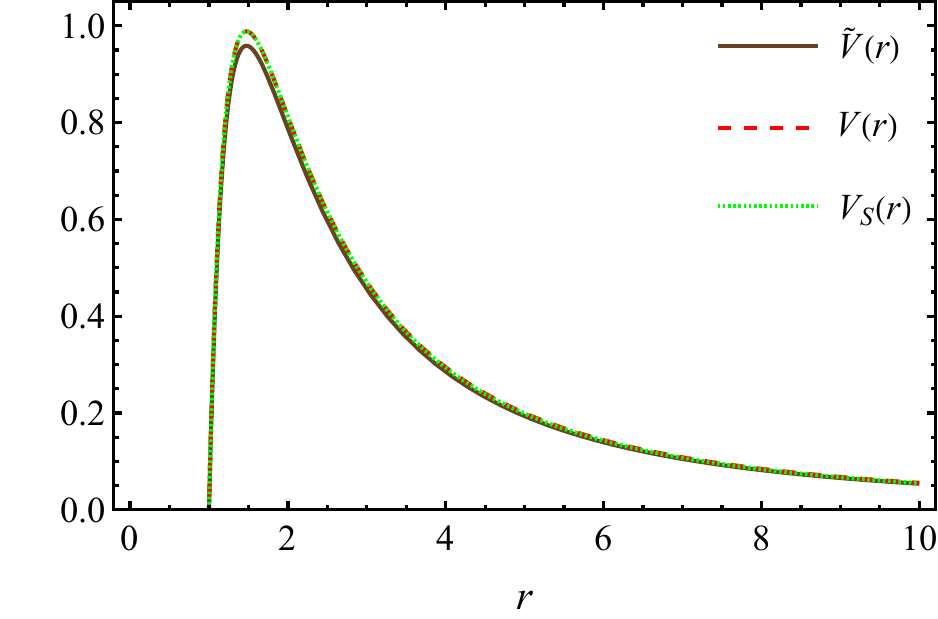}
	\end{center}
	\vspace{-0.7cm}
	\caption{\footnotesize Scalar QNM potential as a function of radial coordinate for $l=2$ and $\gamma_{\text{sp}} =7/3$ for M87* with the mass of $M_\text{BH}=6.4\times 10^9 M_\odot$ surrounded by a DM spike.  $V(r)$, in dashed red, is the potential corresponding to $t$ and $\tilde{V}(r)$, in solid brown, is the potential viewed by an asymptotic observer. We choose  $\rho_{\text{sp}}=1.8 \times 10^{-22}$ g cm$^{-3}$ ($8.4$ times the expected value), $r_\text{b}=2 r_{\text{BH}}$ and $R_{\text{sp}}=4.26$ kpc.  With these values, $\mathcal{C}=1.031$. For comparison, we include the Schwarzschild potential, $V_S(r)$, in dotted green for $M_\text{BH}=6.4\times 10^9 M_\odot$. All our variables are expressed in terms of black hole parameters.}
	\label{fig-M87-potential-8p4}
\end{figure}
\begin{figure}[th!]
	\begin{center}
		\includegraphics[height=5.cm]{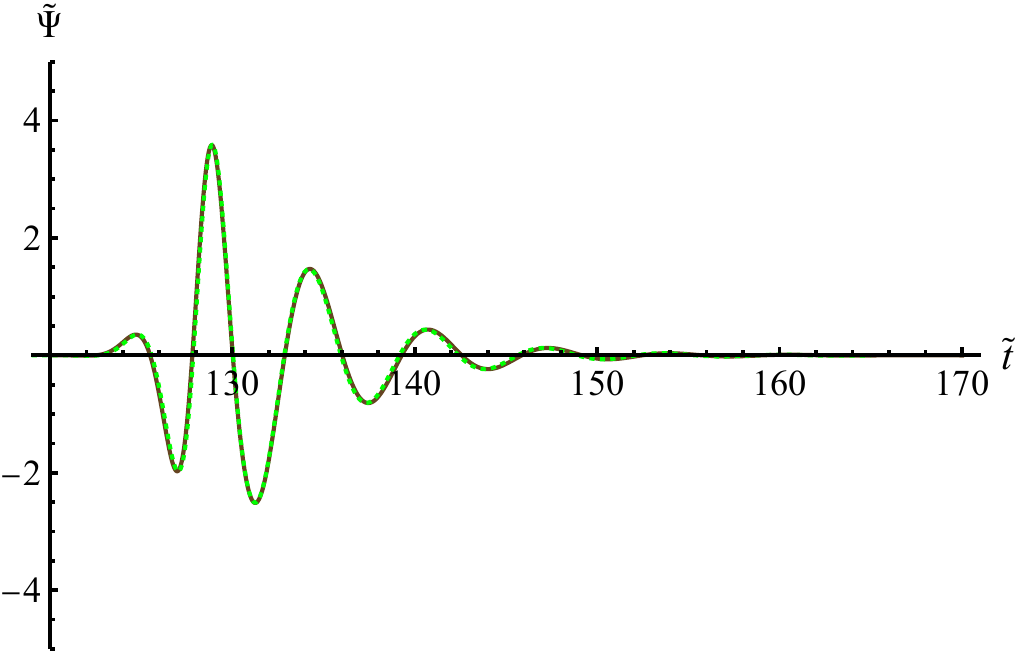}
		\includegraphics[height=5.cm]{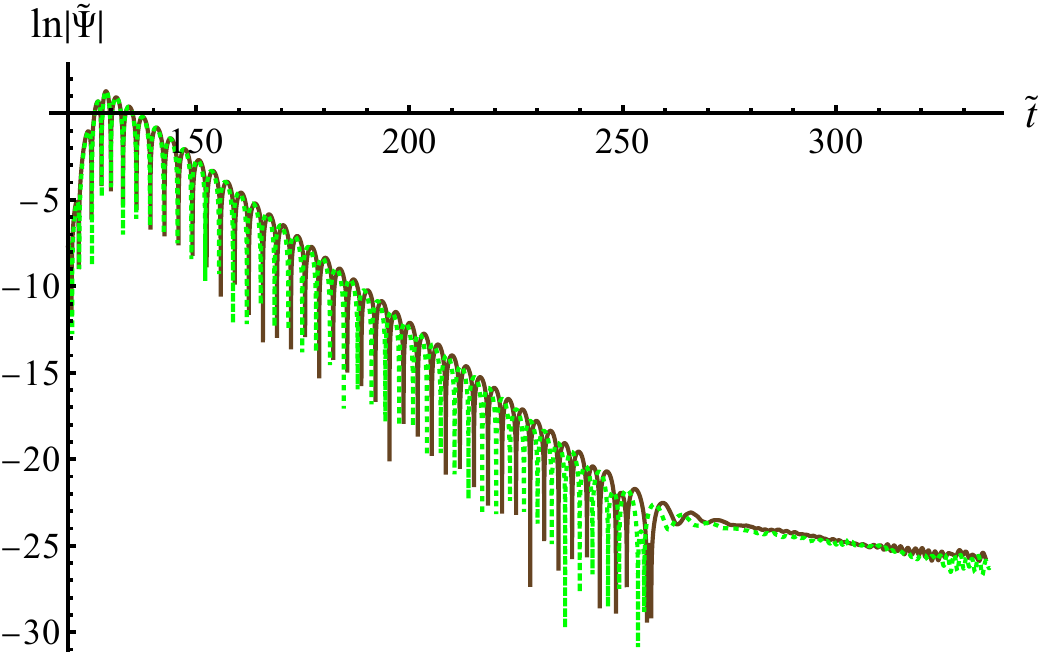}
	\end{center}
	\vspace{-0.7cm}
	\caption{\footnotesize In solid brown, ringdown waveform $\tilde{\Psi}$ (left) and $\ln |\tilde{\Psi}|$ (right) as a function of time $\tilde t$ for the potential, $\tilde{V}(r)$, shown in solid brown in Figure \ref{fig-M87-potential-8p4}.  For numerical purposes, we detect the waveform at $r=80 ~ r_{\text{BH}}$.  For comparison, we include the Schwarzschild ringdown waveform in dotted green. All our variables are expressed in terms of black hole parameters.}
	\label{fig-M87wave-8p4}
\end{figure}
\begin{figure}[th!]
	\begin{center}
		\includegraphics[height=5.3cm]{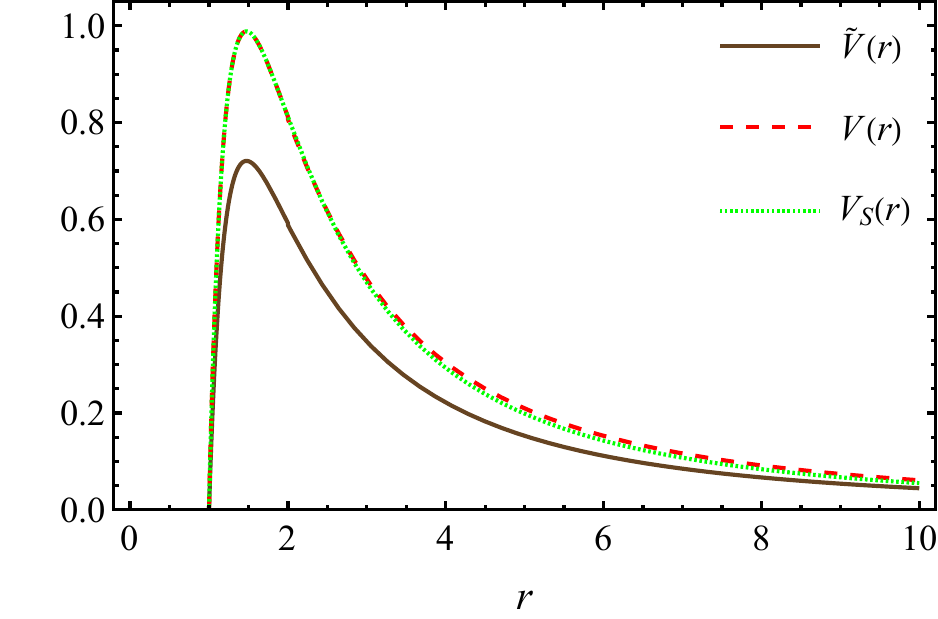}
	\end{center}
	\vspace{-0.7cm}
	\caption{\footnotesize Scalar QNM potential as a function of radial coordinate for $l=2$ and $\gamma_{\text{sp}} =7/3$ for M87* with the mass of $M_\text{BH}=6.4\times 10^9 M_\odot$ surrounded by a DM spike.  $V(r)$, in dashed red, is the potential corresponding to $t$, and $\tilde{V}(r)$, in solid brown, is the potential viewed by an asymptotic observer. We choose  $\rho_{\text{sp}}=1.8 \times 10^{-21}$ g cm$^{-3}$ ($84$ times the expected value), $r_\text{b}=2 r_{\text{BH}}$ and $R_{\text{sp}}=4.26$ kpc.  With these values, $\mathcal{C}=1.371$. For comparison, we include the Schwarzschild potential, $V_S(r)$, in dotted green for $M_\text{BH}=6.4\times 10^9 M_\odot$. All our variables are expressed in terms of black hole parameters.}
	\label{fig-M87-potential-84}
\end{figure}

To see how the DM spike influences the shape of the QNM potential given in Eq.\ (\ref{eq-scalarV}), in Figures \ref{fig-MW-potential}, \ref{fig-M87-potential-8p4}, and \ref{fig-M87-potential-84} we plot the Schwarzschild potential together with the potentials corresponding to the time coordinates $t$ and $\tilde t$, respectively.    Figure \ref{fig-MW-potential} shows the potential for Sgr A* with a DM density at the outer edge of the spike, $\rho_\text{sp}$, approximately $840$ times the expected value given in Table $1$ of \cite{DK-spike} for the case of $\gamma_\text{sp}=7/3$. Note that while the potential for the coordinate $t$  closely follows the Schwarzschild case, the potential for the asymptotic observer is  very different. Figures \ref{fig-M87-potential-8p4} and \ref{fig-M87-potential-84} show the potential for M87* with $\rho_\text{sp} \approx 8.4$ and $\rho_\text{sp} \approx 84$ times the expected value,  provided in Table $2$ of \cite{DK-spike} for the case of $\alpha_\gamma=1.94$, respectively. 
 
\begin{figure}[th!]
	\begin{center}
		\includegraphics[height=5.cm]{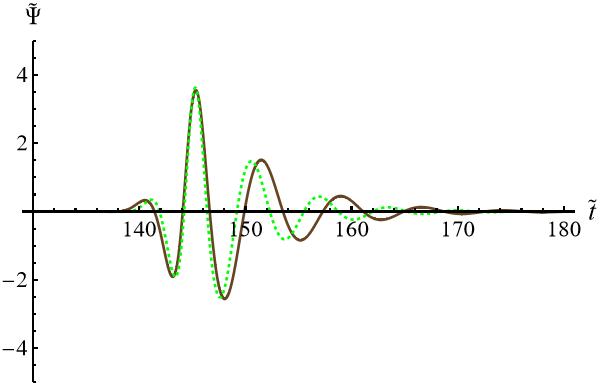}
		\includegraphics[height=5.cm]{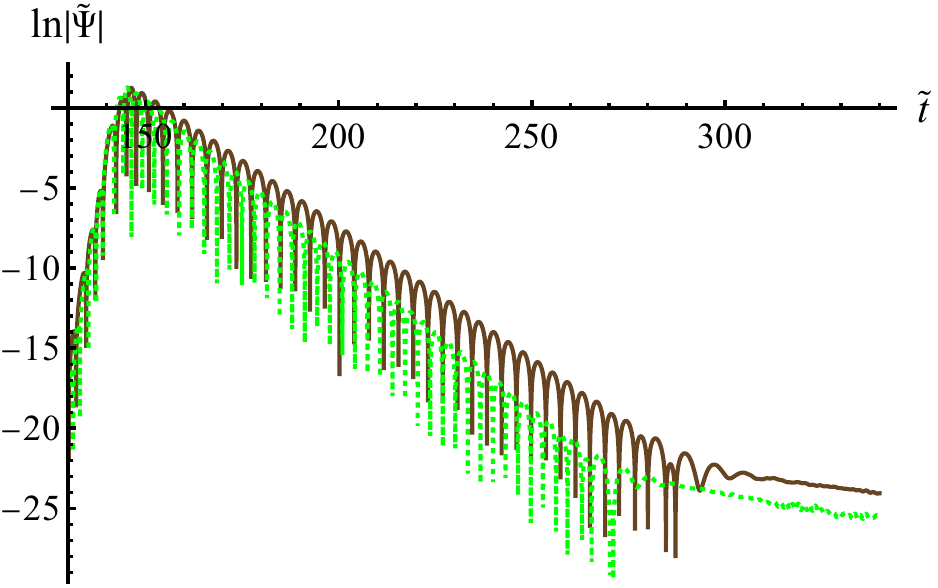}
	\end{center}
	\vspace{-0.7cm}
	\caption{\footnotesize 
	In solid brown, ringdown waveform $\tilde{\Psi}$ (left) and $\ln |\tilde{\Psi}|$ (right) as a function of time $\tilde t$ for the potential, $\tilde{V}(r)$, shown in solid brown in Figure \ref{fig-M87-potential-84}.  For numerical purposes, we detect the waveform at $r=70 ~ r_{\text{BH}}$.  For comparison, we include the Schwarzschild ringdown waveform in dotted green. All our variables are expressed in terms of black hole parameters.}
	\label{fig-M87wave-84}
\end{figure}

The ringdown waveforms for the potentials in Figures \ref{fig-MW-potential}, \ref{fig-M87-potential-8p4}, and \ref{fig-M87-potential-84} are shown in Figures \ref{fig-MWwave}, \ref{fig-M87wave-8p4}, and \ref{fig-M87wave-84} respectively.  These waveforms are  caused by the outgoing Gaussian pulse used in \cite{DK-spike} that initiates near the event horizon (inside the photon sphere) and gets detected after being transmitted through the QNM potential.  In these figures, we also provide the ringdown waveform for the Schwarzschild case for comparison. Note that for the more massive black hole M87*, less density at the outer edge of the spike is required to achieve an observable signal.   In Figures (\ref{fig-M87-potential-8p4}) and (\ref{fig-M87wave-8p4}), the change in the potential and ringdown waveform   due to the presence of DM spike is visible in the asymptotic frame even for a density as low as $\approx 8.4$ times the expected $\rho_\text{sp}$.  This is not the case for an observer who uses the time coordinate $t$. 
\begin{figure}[th!]
	\begin{center}
		\includegraphics[height=5.cm]{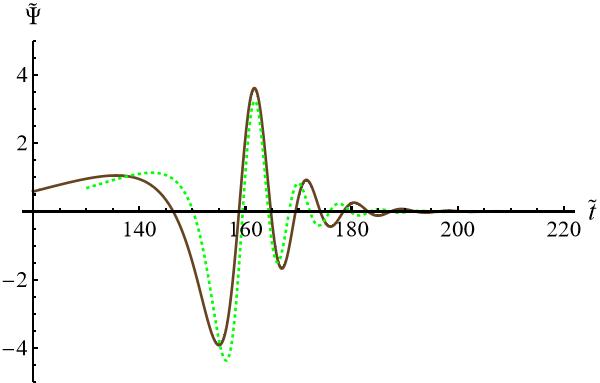}
		\includegraphics[height=5.cm]{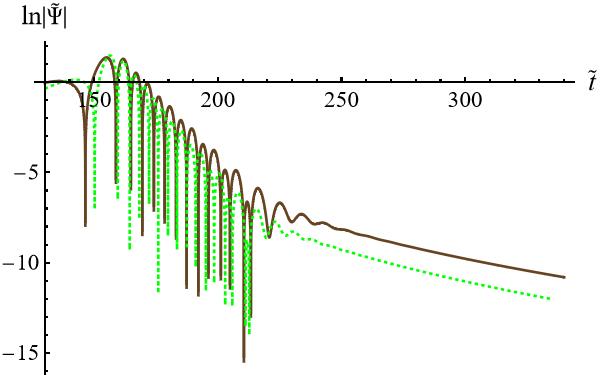}
	\end{center}
	\vspace{-0.7cm}
	\caption{\footnotesize	
	In solid brown, ringdown waveform $\tilde{\Psi}$ (left) and $\ln |\tilde{\Psi}|$ (right) as a function of time $\tilde t$ for the potential, $\tilde{V}(r)$, shown in solid brown in Figure \ref{fig-M87-potential-84}.  Unlike all the other figures, the ringdown is produced by an ingoing pulse starting at $r\approx 41 ~ r_{\text{BH}}$.  For numerical purposes, we detect the waveform at $r=70 ~ r_{\text{BH}}$.  For comparison, we include the Schwarzschild ringdown waveform in dotted green. All our variables are expressed in terms of black hole parameters.}
	\label{fig-M87wave-84-out}
\end{figure}

Finally,  in Figure \ref{fig-M87wave-84-out}, we generate the ringdown waveform produced by the same Gaussian pulse that moves inward from a far distance outside the potential  $\tilde{V}(r)$ shown in Figure \ref{fig-M87-potential-84} and reflects back to generate the ringdown waveform.  The reflected wave produces considerably fewer oscillations than the transmitted wave shown in Figure \ref{fig-M87wave-84}.  However, the transmitted and reflected waves are consistent in terms of the observed impact of the DM spike.  In both cases, the main contribution to the waveform modification comes from the redshift caused by the extra mass in the system.

Note that  it is not possible numerically to generate these ringdown waveforms for an observer at a region outside the spike, since $R_\text{sp}$ is of the order of at least $10^7$ times the black hole radius.  However, the potential drops very quickly with distance.  For example, at $r=50~ r_\text{BH}$ the potential is three orders of magnitude smaller than its value at the peak.   This allows us to generate the ringdown waveform for an observer located  at a radius close to the black hole (roughly around $80~ r_\text{BH}$).  Since the potential is already relatively small at this radius, we do not expect any significant modification in the waveform as it travels all the way to the region outside the spike.

It is evident from the results in this section that the impact of the  DM spike on the ringdown waveform is more pronounced  in the frame of an asymptotic observer, compared to the results presented in \cite{DK-spike}. More specifically, to generate roughly the same modification in the ringdown waveform, we need an order of magnitude less DM density in terms of the proper time $\tilde t$ of an asymptotic observer compared to an observer who uses $t$.  We can, therefore, conclude that if a significant gravitational wave detection associated with perturbations of a black hole with a mass comparable to M87* occurs, it might provide the means to detect the presence of a DM spike or at least put a model dependent bound on its parameters. More massive galactic black holes could likely yield even clearer signals, as argued in \cite{DK-spike}.

\section{Galactic Black Holes with Dark Matter Halos}
\label{Sec:DMH}

In this section, we investigate a black hole surrounded by a DM halo in order to  compare our results with the spike case.  Since the observational signal is more significant for larger black holes, we focus on M87*. The density function of the DM halo in an elliptical galaxy, such as M87, is well-described by the Hernquist profile that has the form \cite{Hernquist}
\be
\rho(r) =\frac{M_\text{halo} a_0}{2 \pi r (r+a_0)^3}~,
\ee
where $M_\text{halo}$ is the total mass of the halo and $a_0$ is the scale radius.  The scale radius of a halo is typically of the order $a_0 \approx 10^4  M_\text{halo}$ \cite{NFW}.   
After integrating the TOV Eq. (\ref{eq:Gtt}) with the above density profile, we obtain 
\be
M(r)=  M_\text{total}-M_\text{halo} a_0\frac{2r+a_0}{(r+a_0)^2}~,
\label{eq-haloM}
\ee
where we choose the constant of integration to be $M_{total}=M_\text{BH}+M_\text{halo}$.

\begin{figure}[th!]
	\begin{center}
		\includegraphics[height=5.1cm]{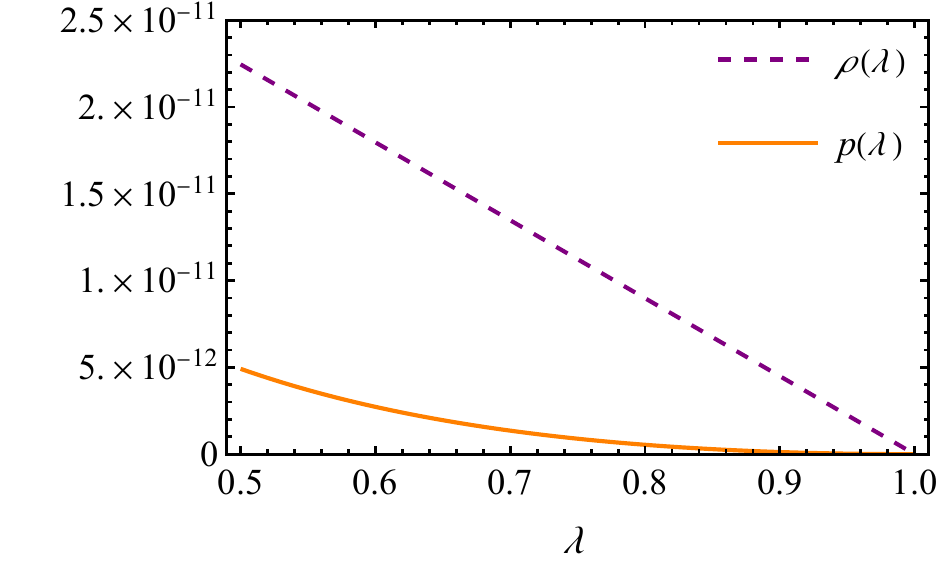}
		\includegraphics[height=5.1cm]{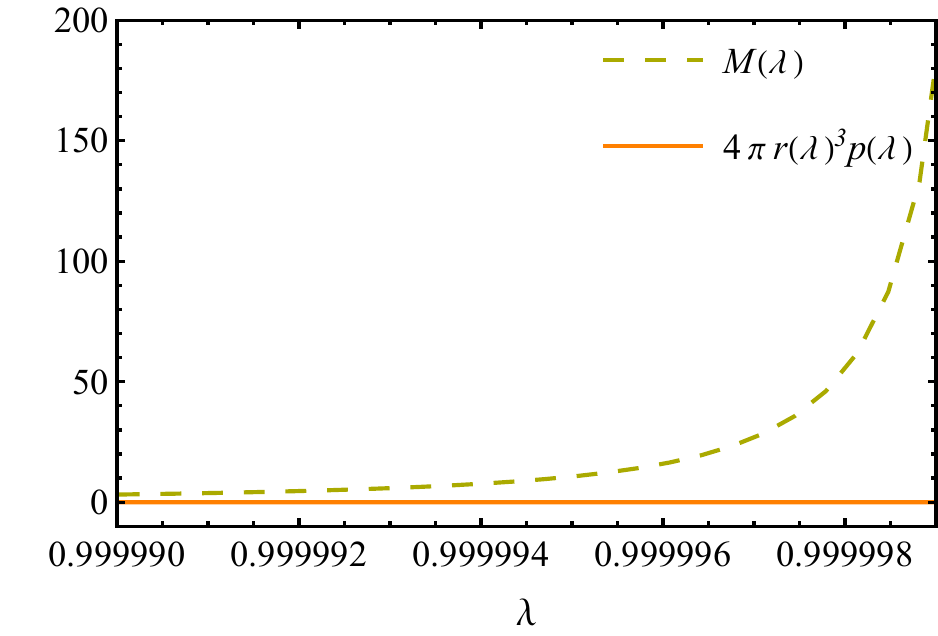}
	\end{center}
	\vspace{-0.7cm}
	\caption{\footnotesize On the left, the DM halo pressure $p$ as a function of  $\lambda=1-2M_\text{BH}/r$ is plotted in solid orange for $a_0=1000 M_\text{halo}$,  where $M_\text{halo}=4.54 \times 10^{13} M_\odot$.  The pressure is  obtained numerically using Eq.\ (\ref{eq:MomentumConservation}). For comparison, we include the DM halo density $\rho(\lambda)$ in dashed purple.    On the right, for the same parameters, we plot the term $4 \pi r(\lambda)^3 p(\lambda)$ and $M(\lambda)$ to show that the pressure term stays negligible while $M(\lambda)$ increases from $M_\text{BH}$ to $M_\text{total}$ as $\lambda$ approaches $1$ ($r\rightarrow \infty$).  All our variables are expressed in terms of black hole parameters.}
	\label{fig-PressureHalo}
\end{figure}

In the anisotropic case with negligiable radial pressure, $p_r$, one can proceed to Eq.\ (\ref{eq:alpha}) below.  In the isotropic case, which is physically less relevant,
we need to solve Eq.\ (\ref{eq:MomentumConservationIsotropic}) for $p(r)$. This is not possible analytically, but we have solved it numerically using the built-in {\em Mathematica} commands for solving differential equations.  It turns out that the pressure term, $4\pi r^3 p(r)$, can be neglected compared to $M(r)$ in the TOV equations. We show the numerical results in Fig.\ \ref{fig-PressureHalo} for the case where $a_0=1000 M_\text{halo}$.  To better manage the numerical calculations, we solve Eq.\ (\ref{eq:MomentumConservation}) in the coordinate $\lambda=1-2M_\text{BH}/r$, which maps the interval $r_\text{BH}\le r < \infty$ to $0\le \lambda <1$.  While pressure and density are comparable in magnitude, it is clear from Fig.\ \ref{fig-PressureHalo} that $4\pi r^3 p(r) << M(r)$.  More specifically, the term  $4\pi r^3 p(r)$ stays close to zero while $M(r)$ increases from $M_\text{BH}$ to $M_\text{total}$ at large distances from the black hole.  

In the absence of pressure, anisotropic and isotropic cases become the same. This is because  for both cases the TOV equation (\ref{eq:Grr})  reduces to Eq.\ (\ref{eq:GrrNonIsotropic}).  Combining Eqs.\ (\ref{eq:GrrNonIsotropic}) and (\ref{eq-haloM}) gives
\bea
\frac{d\mu(r)}{dr} = -\frac{1}{r} +\frac{1}{r-2\left[M_\text{total}-M_\text{halo} a_0\frac{2r+a_0}{(r+a_0)^2}\right]}~.
\label{eq:alpha}
\eea
We  integrate Eq.\ (\ref{eq:alpha}) to get
\begin{eqnarray}
	\mu(r) &=&-\int  \frac{dr}{r} +\int \frac{y^2 dy}{y^3-(2 M_\text{total}+a_0) y^2 + 4 M_\text{halo} a_0 y -2 M_\text{halo} a_0^2 }+C \nonumber \\
	&=& -\int  \frac{dr}{r} 
	+\frac{ y_0^2}{y_1 y_2+y_0(y_0-y_1-y_2)}\int \frac{ dy}{y-y_0}\nonumber \\
	&& +\frac{1 }{y_1 y_2+y_0(y_0-y_1-y_2)}\int \frac{(y_0 y_1 y_2-y_0 y_1 y-y_0 y_2 y+y_1 y_2 y) dy}{y^2-(y_1+y_2)y+y_1 y_2}+C,
	\label{eq: alpha7:3}
\end{eqnarray}
where we have used the change of variable $y=r+a_0$.  Here, $y_0$ is the real root of the equation $y^3-(2 M_\text{total}+a_0) y^2 + 4 M_\text{halo} a_0 y -2 M_\text{halo} a_0^2$ and $y_1$ and $y_2$ are the two complex conjugate roots. After integration, the final result for the metric function, $A(r)=e^{\mu(r)}$, is
\begin{eqnarray}
	A(r) &=&\left(1-\frac{2M_{\text{BH}}}{r_\text{in}}\right)\frac{r_\text{in}}{r} \left(\frac{r+a_0-y_0}{r_\text{in}+a_0-y_0}\right)^{\frac{y_0^2}{y_1 y_2+y_0(y_0-y_1-y_2)}} \nonumber \\
	&&\left(\frac{(r+a_0)^2-(y_1+y_2)(r+a_0)+y_1y_2}{(r_\text{in}+a_0)^2-(y_1+y_2)(r_\text{in}+a_0)+y_1y_2}\right)^{\frac{[y_1 y_2-y_0(y_1+y_2)]}{2[y_1 y_2+y_0(y_0-y_1-y_2)]}} \nonumber \\
	&&e^{   \frac{2 y_0 y_1 y_2-y_0(y_1+y_2)^2+y_1 y_2 (y_1+y_2)}{\sqrt{4y_1y_2-(y_1+y_2)^2}[y_1 y_2+y_0(y_0-y_1-y_2)]} \left(\arctan{\frac{2(r+a_0)-y_1-y_2}{\sqrt{4y_1y_2-(y_1+y_2)^2}}}-\arctan{\frac{2(r_\text{in}+a_0)-y_1-y_2}{\sqrt{4y_1y_2-(y_1+y_2)^2}}}\right)}.
	\label{eq:A-halo}
\end{eqnarray}
We have chosen the constant of integration, $C$, so that $A(r_\text{in}) =1-2M_{\text{BH}}/r_\text{in}$.

\begin{figure}[th!]
	\begin{center}
		\includegraphics[height=5.9cm]{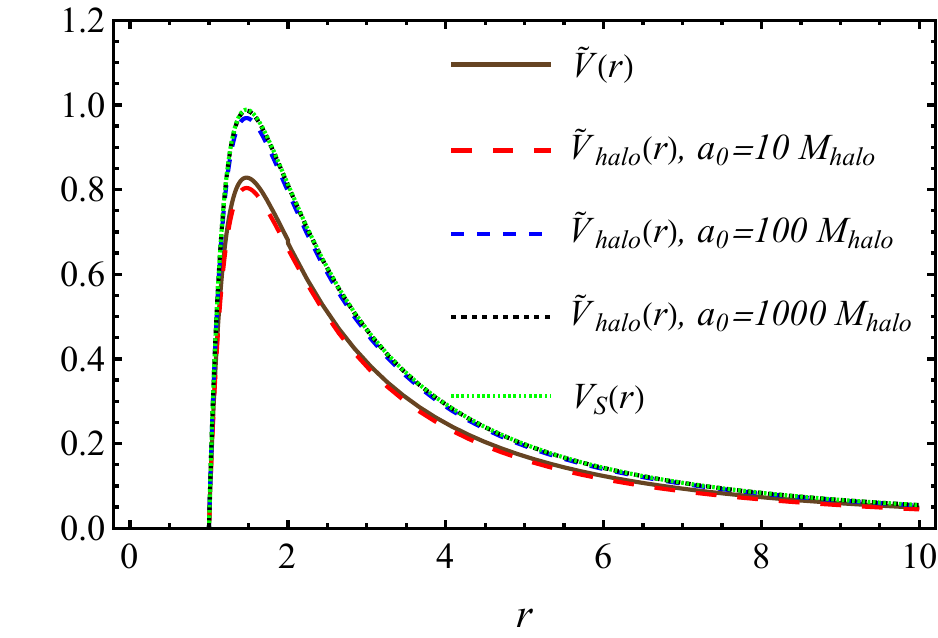}
	\end{center}
	\vspace{-0.7cm}
	\caption{\footnotesize Scalar QNM potential as a function of radial coordinate for $l=2$ for M87* with the mass of $M_\text{BH}=6.4\times 10^9 M_\odot$ surrounded by a DM halo with the mass of $M_\text{halo}=4.54 \times 10^{13} M_\odot$.  The halo extends from $r_\text{in}=2 r_{\text{BH}}$ to infinity.  $\tilde{V}_\text{halo}(r)$ is plotted for $a_0=10 M_\text{halo}$, $a_0= 100 M_\text{halo}$, and $a_0= 1000 M_\text{halo}$ in dashed red, dashed blue, and dashed black respectively.   For comparison, we also plot $\tilde{V}(r)$ in solid brown for a DM spike with the same mass, which extends between $r_\text{in}=2 r_{\text{BH}}$ and $r_{\text{out}}=4.26$ kpc.  We choose the same $r_{\text{out}}$ for the DM halo to calculate the redshift factor.  This assumes the observer is located at $r=r_\text{out}$.  Choosing $r_{\text{out}}\sim \infty$ changes the results by only $0.1\%$.
		In addition, we include the Schwarzschild potential, $V_S(r)$, in dotted green for $M_\text{BH}=6.4\times 10^9 M_\odot$. All our variables are expressed in terms of black hole parameters.}
	\label{fig-M87-potential-100}
\end{figure}

Using Eqs.\ (\ref{eq:b}), (\ref{eq-haloM}), and (\ref{eq:A-halo}) we can obtain the scalar QNM potential (\ref{eq-scalar-tildeV}) for the DM halo.  We show the scalar QNM potential, $\tilde{V}_\text{halo}(r)$, for different values of $a_0$ in Fig.\ \ref{fig-M87-potential-100}.   The QNM potential of the halo is only comparable to the spike of the same mass for $a_0=10 M_\text{halo}$.  In this case,  approximately $99\%$ of the halo mass is contained in the spike region of $r_\text{in} \le r < r_\text{out}$.  For larger values of $a_0$, the mass of the halo spreads out further from the black hole and the QNM potential becomes almost indistinguishable from the Schwarzschild case (balck hole with no DM) for $a_0 > 100  M_\text{halo}$.     In other words, the impact of  the redshift on the QNM potential is negligible for a halo as long as its distribution scale is within a few orders of magnitude of the value expected from observations.  This indicates  the observational signal does depend on how DM is distributed around a black hole. 

\begin{figure}[th!]
	\begin{center}
		\includegraphics[height=6.cm]{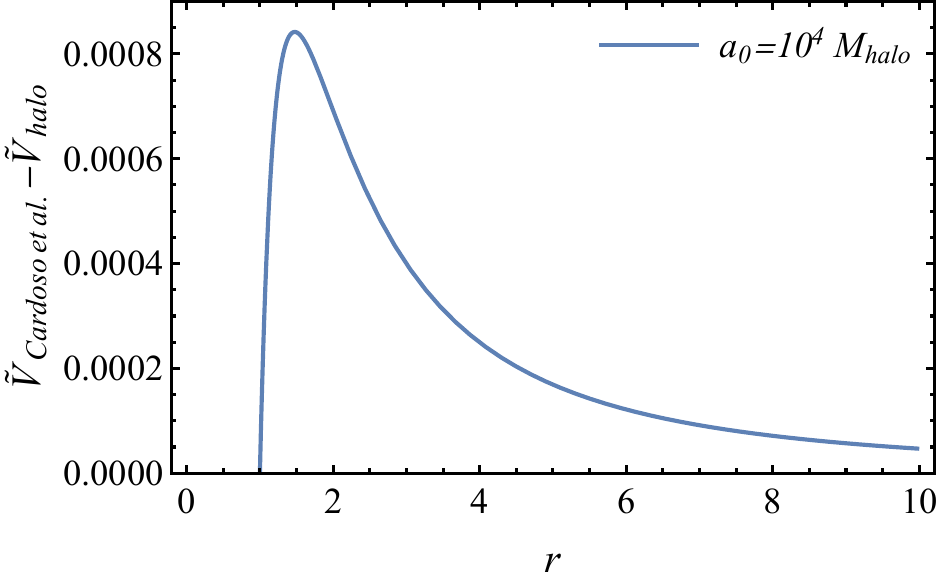}
	\end{center}
	\vspace{-0.7cm}
	\caption{\footnotesize The difference between the scalar QNM potential for $l=2$  obtained using the metric functions derived in this paper and the DM halo model suggested by Cardoso {\it et al.}\cite{CardosoHalo} is shown for $M_\text{BH}=6.4\times 10^9 M_\odot$ and $M_\text{halo}=4.54 \times 10^{13} M_\odot$.
		}
	\label{fig-Cardoso-DifferencePotential}
\end{figure}

In \cite{CardosoHalo}, Cardoso {\it et al.~}have also found an analytic solution for a Hernquist-like density profile for an anisotropic DM halo, with negligible $p_r$, surrounding a black hole.  
In their solution, Cardoso {\it et al.~}modify the Hernquist density function in order to impose the boundary condition  $g_{tt}(r\rightarrow \infty)\rightarrow 1$.  The solution of \cite{CardosoHalo} turns out to be a close match to the  solution we find here, where we do not alter the Hernquist density function.  
In Figure \ref{fig-Cardoso-DifferencePotential}, we show the difference between the scalar QNM potential obtained using our metric functions (\ref{eq-haloM}) and (\ref{eq:A-halo}) and the potential found using the metric functions provided in \cite{CardosoHalo}. The difference between the two solutions is less than $0.1\%$.

\section{Shadow Radius}
\label{Sec:Shadow}

We summarize the shadow calculations as follows.\footnote{For a good review of black hole shadow calculations, see \cite{Shadow-rev}.}  Consider the action for particle geodesics
\begin{equation}
	S =  \int_{\lambda_0}^{\lambda_f} L(x,\dot{x}) d\lambda = \frac{1}{2} \int_{\lambda_0}^{\lambda_f}  g_{\mu \nu} \dot{x}^\nu \dot{x}^\mu  d\lambda  ,
	\label{}
\end{equation}
where $L$ is the Lagrangian and dot represents derivative with respect to the affine parameter $\lambda$.
For the general static spherically symmetric spacetime in Eq.\ (\ref{eq:GeneralMetric}), the Lagrangian is
\begin{equation}
	L(x, \dot{x})= \frac{1}{2}\left[-A(r) \dot{t}^2 + B(r)^{-1} \dot{r}^2 +r^2 (\dot{\theta}^2+\sin^2\theta \dot{\phi}^2)\right].
	\label{}
\end{equation}
The spherical symmetry allows us to take $\theta=\pi/2$ without the loss of generality.  The $t$ and $\phi$ components of the Euler-Lagrange equation
\begin{equation}
	\frac{d}{d\lambda} \left(\frac{\partial L}{\partial \dot{x}^\mu} \right) - \frac{\partial L}{\partial x^\mu}=0
	\label{}
\end{equation}
give us two constants of motion $E=A(r) \dot{t}$ and  $L=r^2 \dot{\phi}$, which correspond to energy and angular momentum respectively.

For null geodesics we have $g_{\mu \nu} \dot{x}^\mu \dot{x}^\nu =0$.  Hence
\begin{equation}
	-A(r) \dot{t}^2 + B(r)^{-1} \dot{r}^2 +r^2 \dot{\phi}^2=0.
	\label{}
\end{equation}
After combining the constants of motion with the above equation and using the fact that $\dot{r}^2/\dot{\phi}^2=(dr/d\phi)^2$, one obtains the orbit equation for null geodesics,
\begin{equation}
	\left( \frac{dr}{d\phi}\right)^2 =r^2 B(r)\left(\frac{r^2}{A(r)} \frac{1}{b^2} -1\right),
	\label{eq: geodesic}
\end{equation}
where $b=L/E$ is the impact parameter defined as the perpendicular distance, measured at infinity, between the geodesic and the parallel line that passes through the center of black hole in an asymptotically Minkowski spacetime\footnote{Specifically, the spacetime must be asymptotically flat and the metric must be asymptotically Minkowski so that $A(r)\to 1$ as $r\to \infty$.} .  The  closest distance, $R$, of the light ray to the black hole is the turning point of the geodesic, where the condition $dr/d\phi|_R=0$ has to hold.  This condition gives us a relation between $R$ and $b$,
\begin{equation}
	b^2=\frac{R^2}{A(R)}.
	\label{eq:ImpactParameter}
\end{equation}
When $R=r_\text{ph}$, where $r_\text{ph}$ is the radius of the photon sphere, the light ray that is coming from infinity will orbit around the black hole.  In this situation, in addition to $dr/d\phi|_{r_\text{ph}}=0$, the condition $d^2r/d\phi^2|_{r_\text{ph}}=0$ should also hold.  Combining these two conditions with Eq.\ (\ref{eq: geodesic}) gives
\begin{equation}
	\frac{1}{A(r)}\frac{dA(r)}{dr}=\frac{2}{r},
	\label{eq:photon-sphere}
\end{equation}
which can be used to determine the radius of the photon sphere.  It is easy to show that for the Schwarzschild spacetime, where $A(r)=1-2M_\text{BH}/r$, Eq.\  (\ref{eq:photon-sphere}) gives $r_\text{ph}=3M_\text{BH}$.  Once we determine $r_\text{ph}$, we can use Eq.\ (\ref{eq:ImpactParameter}) to find the corresponding impact parameter 
\begin{equation}
	b_\text{ph}=\frac{r_\text{ph}}{\sqrt{A(r_\text{ph})}}~.
\end{equation}
The shadow radius, $r_\text{sh}$, turns out to be equal to $b_\text{ph}$, because the rays with $b < b_\text{ph}$ cannot escape the black hole.  It is important, however, to note that $b$ only corresponds to the impact parameter at infinity if we assume the black hole spacetime is asymptotically Minkowski.   For asymptotically flat black holes,  $r_\text{sh}$ does not need to be equal to $b_\text{ph}$.    This can be explained as the following.  If the distance between the observer and the center of the black hole is $r_\text{o}$ and the angle between $r_\text{o}$ and the geodesic at infinity is $\alpha$, then we can write the relation
\begin{equation}
	\sin \alpha = \frac{\mathcal{I}}{r_\text{o}} ,
	\label{}
\end{equation}
where $\mathcal{I}$ is the impact parameter.  Note that at the observer position, we also have
\begin{equation}
	\sin \alpha \sim \frac{\sqrt{r^2 B(r)}d\phi}{dr}\bigg|_{r_\text{o}}  \sim \left(b^2  \frac{A(r_\text{o})}{r_\text{o}^2} \right)^{1/2}= \left( \frac{R^2}{A(R)}  \frac{A(r_\text{o})}{r_\text{o}^2} \right)^{1/2},
	\label{}
\end{equation}
where we use Eq.\ (\ref{eq: geodesic}) and the small angle approximation.
The above two equations give
\begin{equation}
	\mathcal{I}\sim b  \sqrt{A(r_\text{o}) }.
	\label{}
\end{equation}
It is clear  that $\mathcal{I}=b$ only when $A(r_\text{o})=1$.  Finally, the black hole shadow size is given by
\begin{equation}
	r_\text{sh}=\mathcal{I}_\text{ph} \sim b_\text{ph}  \sqrt{A(r_\text{o}) }= r_\text{ph} \sqrt{\frac{A(r_\text{o})}{A(r_\text{ph})} }.
	\label{}
\end{equation}
Note that the shadow radius is invariant under time rescalings, so that one can obtain the correct result using either $\tilde t$ or $t$.

We now explore how the shadow of a black hole changes in the presence of a DM spike.  In an asymptotically Minkowski space,  $A(r_\text{o}) \approx 1$. For the Schwarzschild case, we find the radius of the shadow to be  
\begin{equation}
	r_\text{sh}^\text{Schw} \sim  r_\text{ph} \sqrt{\frac{1}{A(r_\text{ph})} }=3\sqrt{3}M_\text{BH},
	\label{}
\end{equation}
where we use the fact that $r_\text{ph}=3 M_\text{BH}$.  In the presence of a DM spike or halo, the asymptotic observer will measure the radius to be

\begin{equation}
	r_\text{sh}^\text{sp} \sim  r_\text{ph} \sqrt{\frac{1}{\tilde{A}(r_\text{ph})} }=  r_\text{ph} \sqrt{\frac{1}{A(r_\text{ph})/\mathcal{C}} }=3\sqrt{3\mathcal{C}}M_\text{BH} .
\label{}
\end{equation}
Therefore the black hole shadow radius will appear to be $\sqrt{\mathcal{C}}$ times larger due to the presence of DM.   For the shadow radius of M87*,   bounds on the DM spike density produce an effect\footnote{$\mathcal{C}\approx1.004$ for the M87* spike profile provided in Table 2 of \cite{DK-spike} for the case of $\alpha_\gamma=1.94$.  This leads to an approximately $0.2\%$ enlargement of the shadow diameter. } that is an order of magnitude smaller than is accessible with current Event Horizon Telescope data \cite{EHT-M87-PRIMO}.  In the case of halo, however, the enlargement of the shadow size is negligible as long as the distribution scale of the halo is within a few orders of magnitude of the value expected from observations (see Fig.\ \ref{fig-M87-potential-100}).


\section{Conclusion}
\label{Sec:Conclusion}

We have shown that a DM spike or halo surrounding the black hole at the center of M87  will effect the associated ringdown waveform predominantly in the form of an overall redshift between the frame of an observer who uses the Schwarzschild time below the inner radius of the spike/halo and that of an asymptotic observer. The impact of  the redshift on the asymptotic ringdown waveform is significant in case of the spike, but not the halo as long as the distribution scale of the latter is within a few orders of magnitude of the value expected from observations.  
 We conclude that if it becomes possible to detect the gravitational waves  associated with perturbations of a galactic black hole with a mass comparable or larger than that of M87*, it could provide the means to detect the presence of a DM spike or  put a model dependent bound on its parameters.

We also showed that the presence of a DM spike or halo would increase the black hole shadow radius by the redshift factor of $\sqrt{\mathcal{C}}$.  Remarkably, known bounds on the DM spike density produce an effect  on the shadow radius of M87* that is just an order of magnitude smaller than is accessible with current Event Horizon Telescope data \cite{EHT-M87-PRIMO}. Spikes surrounding more massive black holes would produce even larger effects\cite{DK-spike}  on the shadow and may soon provide the observational means to confirm or rule out their presence.

\vspace{1cm}
\leftline{\bf Acknowledgments}
We thank Andrei Frolov and Don Page for helpful discussions.  G.K.\ gratefully acknowledges that this research was supported in part by a Discovery Grant from the Natural Sciences and Engineering Research Council of
Canada.

\appendix
\section{Discontinuity in Potential}
\label{Sec:Appendix}
 Here, we derive the discontinuities in the QNM potential that  exist at the boundaries of the matter shell.  To do this, we first rewrite Eq.\ (\ref{eq-scalarV}) as
\bea
\frac{V(r)}{A(r)}&=& \frac{l(l+1)}{ r^2}+ \frac{1}{2r} \left[B(r) \frac{A'(r)}{A(r)} +B'(r)\right]\\
&=&\frac{l(l+1)}{r^2}+\frac{1}{2r} \left[\left(1-\frac{2M(r)}{r}\right) \mu'(r) + \left(\frac{-2M'(r)}{r} +\frac{2M(r)}{r^2}\right)\right]~.
\label{eq-scalarVb}
\eea
Using the TOV equations (\ref{eq:Gtt}) and (\ref{eq:Grr}), we can write
\bea
\frac{V(r)}{A(r)}
&=&\frac{l(l+1)}{r^2}+\frac{1}{2r} \left[\left(1-\frac{2M(r)}{r}\right)\left[\frac{2M(r)}{r(r-2M(r))}\right] + \left(\frac{- 8\pi r^2 \rho(r)}{r} +\frac{2M(r)}{r^2}\right)\right]\nn\\
&=&\frac{l(l+1)}{ r^2}+\frac{1}{2r^2} \left[\frac{2M(r)}{r} + \left({- 8\pi r^2 \rho(r)} +\frac{2M(r)}{r}\right)\right]\nn\\
&=&\frac{l(l+1)}{ r^2}+\frac{1}{r^3} \left[2M(r) - 4\pi r^3\rho(r) \right]~.
\label{eq-scalarVb}
\eea
Therefore,
\bea
V(r) = A(r)\left[\frac{l(l+1)}{ r^2}+\frac{1}{r^3} \left(2M(r) - 4\pi r^3\rho(r) \right)\right]
\eea
and since only $\rho$ is discontinuous, the discontinuity in $V(r)$ is
\bea
\Delta V(r) &=& -4\pi A(r) \Delta \rho(r)~.
\eea
Thus, the discontinuities in the QNM potential are proportional to the change in matter density at the boundaries, and consequently negligibly small for proposed DM spike profiles \cite{Nampalliwar, DK-spike, M87data}.  



\def\jnl#1#2#3#4{{#1}{\bf #2} #3 (#4)}

\def\Zphys{{Z.\ Phys.} }
\def\jssc{{J.\ Solid State Chem.\ }}
\def\jpsJ{{J.\ Phys.\ Soc.\ Japan }}
\def\ptps{{Prog.\ Theoret.\ Phys.\ Suppl.\ }}
\def\PTP{{Prog.\ Theoret.\ Phys.\  }}
\def\LNC{{Lett.\ Nuovo.\ Cim.\  }}

\def\JMP{{J. Math.\ Phys.} }
\def\NPB{{Nucl.\ Phys.} B}
\def\NP{{Nucl.\ Phys.} }
\def\PLB{{Phys.\ Lett.} B}
\def\PL{{Phys.\ Lett.} }
\def\PRL{Phys.\ Rev.\ Lett.\ }
\def\PRA{{Phys.\ Rev.} A}
\def\PRB{{Phys.\ Rev.} B}
\def\PRD{{Phys.\ Rev.} D~}
\def\PR{{Phys.\ Rev.} }
\def\PRe{{Phys.\ Rep.} }
\def\AP{{Ann.\ Phys.\ (N.Y.)} }
\def\RMP{{Rev.\ Mod.\ Phys.} }
\def\ZPC{{Z.\ Phys.} C}
\def\SCI{Science}
\def\CMP{Comm.\ Math.\ Phys. }
\def\MPLA{{Mod.\ Phys.\ Lett.} A}
\def\IJMPA{{Int.\ J.\ Mod.\ Phys.} A}
\def\IJMPB{{Int.\ J.\ Mod.\ Phys.} B}
\def\cmp{{Com.\ Math.\ Phys.}}
\def\JPA{{J.\  Phys.} A}
\def\CQG{Class.\ Quant.\ Grav.~}
\def\ATMP{Adv.\ Theoret.\ Math.\ Phys.~}
\def\AJP{Am.\ J.\ Phys.~}
\def\PRSA{{Proc.\ Roy.\ Soc.\ Lond.} A }
\def\ibid{{ibid.} }
\vskip 1cm


\end{document}